\documentclass[onecolumn]{mn2e}
\usepackage{times}
\input{psfig.sty}
\newif\ifAMStwofonts

\begin{document}

\title[SN 1993J: a veiled pulsar in a binary system?]{Supernova 1993J:
a veiled pulsar in a binary system?}

\author[Ramirez-Ruiz \& Serenelli]{Enrico Ramirez-Ruiz$^{1}$ and Aldo
M. Serenelli \\ School of Natural Sciences, Institute for Advanced
Study, Einstein Drive, Princeton, NJ 08540, USA.\\ $^{1}$Chandra
Fellow.}

\date{}

\maketitle

\label{firstpage}

\begin{abstract}
The recent report of a binary companion to the progenitor of supernova
1993J may provide important clues for identifying the nature of the
nascent compact object. Given the estimates of the progenitor mass,
the potential power source is probably a pulsar rather than an
accreting black hole. If there is a pulsar, one would expect the
rotational luminosity to be stored and reprocessed by the supernova
remnant, but no pulsar nebula has yet been seen. The lack of detection
of an X-ray synchrotron nebula should be taken as strong evidence
against the presence of a bright pulsar. This is because absorption by
the surrounding supernova gas should be negligible for the light of
the companion star to have been detected. A model is developed here
for the luminosity of the pulsar nebula in SN 1993J, which is then
used to predict the spin-down power of the putative pulsar. If one
exists, it can be providing no more than $\sim 6 \times 10^{39} {\rm
erg\;s^{-1}}$. With an initial rotation period of 10-30 ms, as
extrapolated from young galactic pulsars, the nascent neutron star can
have either a weak magnetic field, $B \le 10^{11}$ G, or one so
strong, $B \ge 10^{15}$ G, that its spin was rapidly slowed down. The
companion star, if bound to the neutron star, should provide ample
targets for the pulsar wind to interact and produce high-energy
gamma-rays. The expected non-pulsed, GeV signal is calculated; it
could be detected by current and future experiments provided that the
pulsar wind velocity is similar to that of the Crab Nebula.
\end{abstract}

\begin{keywords}
{pulsars: general -- stars: supernovae -- supernovae:
individual (SN 1993J) -- gamma rays: theory -- radiation
mechanisms:non-thermal}
\end{keywords}

\section{Introduction} 
Supernova 1993J was discovered on 28 March 1993 in the spiral galaxy
M81 (at a distance of 3.63 Mpc; Freedman et al. 1994). Although it is
factor of 65 further away than SN 1987A, it is relatively close for a
SN and has been the subject of intense observational campaigns in a
number of wavelengths regions (Chevalier 1997, and references
therein). Shortly after the explosion, its spectrum was found to
contain hydrogen lines - a type II supernova - but within a few weeks
strong helium lines developed. The spectrum looked then more like a
type Ib supernova, as though the envelope of hydrogen had lifted to
reveal a helium layer (Filippenko et al. 1993). This, together with the
observed peculiarities observed in the lightcurve, led many to
conclude that the dying star, identified as a non-variable red
supergiant in images taken before the explosion (Aldering et
al. 1994), must have lost a considerable amount of its outer envelope
of hydrogen gas to a companion before it exploded (Nomoto et al. 1993;
Podsiadlowski et al. 1993; Woosley et al. 1994). But no companion had
been seen until recently when the brightness of SN 1993J dimmed
sufficiently that its spectrum showed the features of a massive star
superimposed on the supernova (Maund et al. 2004).

Given the estimates of the progenitor mass, the remnant is probably a
neutron star rather than a black hole (Nomoto et al. 1993;
Podsiadlowski et al. 1993; Woosley et al. 1994), and the neutron star
is generally expected to manifest as a pulsar. The newly born neutron
star is characterized by its initial rotation rate and magnetic dipole
moment. It will spin down, generating radiation and accelerating
charged particles at the expense of its rotational energy.  We expect
the reprocessing of rotational energy to produce a bright
flat-spectrum synchrotron nebula, but this soft emission has not been
observed.

Here we developed a simple shocked wind model for the luminosity of
the pulsar nebulae in SN 1993J (Section 2), which is then compared
with the current luminosity limits in order to place bounds on the
dipole emission of the nascent neutron star and its birth properties
(Section 4). We also investigated the impact of the soft photon field
radiation of the binary companion on the central pulsar wind (Section
3), along with the types of observation that would help to
unambiguously demonstrate whether or not the system is disrupted as a
result of the supernova. Our conclusions are discussed in Section 4.

\section{A Model for the Pulsar Nebula in SN 1993J}
Neutron stars formed in core collapse SN explosions are thought to
emit radiation as pulsars, the Crab (Rickett \& Seiradakis 1982) being
the canonical example. Pulsars are generally presumed to lose
rotational energy by some combination of winds of relativistic
particles and magnetic dipole radiation at the frequency of the
neutron star's rotation.  In the generic pulsar model, the field is
assumed to maintain a steady value, and the luminosity declines as the
spin rate slows down (Shapiro \& Teukolsky 1983). The spindown law is
then given by $\Omega(t)=\Omega_0 (1+ t/t_\Omega)^{-1/2}$, so that
$L_\Omega (t) =L_{\Omega,0}(1+ t/t_\Omega)^{-2}$, where $t_\Omega
\simeq 10^3~{\rm s}~ I_{45}B_{p,15}^{-2}P_{0,-3}^2R_6^{-6}$ is the
characteristic time scale for dipolar spindown,
$B_{p,15}=B_p/(10^{15}{\rm G})$ is the dipolar field strength at the
poles, $R_6=R/10^6{\rm cm}$ is the radius of the light cylinder,
$P_{0,-3}$ is the initial rotation period in milliseconds, and
$L_{\Omega,0} \simeq 10^{49} {\rm erg~s^{-1}} B_{p,15}^2 P_{0,-3}^{-4}
R_6^6$.

The power output of the pulsar is typically assumed to be
$L_{\Omega}$, regardless of the detailed process by which the dipole
emission is converted to the energy of charged particle acceleration
and high-energy electromagnetic radiation. The pulses themselves, even
when seen in the gamma-rays, constitute an insignificant fraction of
the total energy loss. Large fraction of the power output, on the
other hand, goes into a bubble of relativistic particles and magnetic
field surrounding the pulsar (Chevalier \& Fransson 1992). This bubble
gains energy from the pulsar, and loses it in synchrotron radiation
losses and in doing work on the surrounding supernova gas, sweeping it
up and accelerating it.

\subsection{Hydrodynamic Evolution}
The density structure of the supernova gas is determined by the
initial stellar structure, as modified by the explosion (Arnett
1988). The density structure has been relatively well determined for
models of SN 1987A -- it shows an inner flat profile outside of which
is a steep power-law decrease with radius (e.g. Shigeyama \& Nomoto
1990). Chevalier \& Soker (1989) approximated the velocity profile by
an inner section $\rho_i=Ar^{-m}t^{m-3}$ and an outer section
$\rho_0=Br^{-n}t^{n-3}$, where $m=1$ and $n=9$ for SN 1987A. The two
segments of the density profile are assumed to be continuous and they
intersect at a velocity $v_t$. For an explosion with total energy $E$
and mass $M_{\rm t}$, one has
\begin{equation}
v_t=\left[{2(5-m)(n-5)E \over (3-m)(n-3)M_{\rm t}}\right]^{1/2},
\end{equation}
where 
\begin{equation}
A=\left({n-3\over n-m}\right)\left[{(3-m)(n-3)M_{\rm t} \over
2(5-m)(n-5)E}\right]^{(3-m)/2}{(3-m)M_{\rm t} \over 4\pi}.
\end{equation}

SN 1987A was an unusual type II supernova because of its small initial
radius. This is in contrast to SN 1993J where the early lightcurve
indicated that the progenitor star was more extended, about ten times
larger than that of SN 1987A. The lack of a plateau phase in this case
is attributed to the low mass of the hydrogen envelope, about 0.1 to
0.6 $M_\odot$ (Nomoto et al. 1993; Podsiadlowski et al. 1993; Woosley
et al. 1994), although the initial stellar mass was probably $\sim
15M_\odot$. Woosley et al. (1994), for example, estimated that the
final presupernova star\footnote{This model was derived from a
13$M_\odot$ main-sequence star that lost most of its hydrogen envelope
to a nearby companion ($9M_\odot$, initially 4.5 AU).} had a helium
and heavy element core of $\sim 3.71 M_\odot$, a low density hydrogen
envelope of $0.2M_\odot$, and a radius of about $4.3 \times 10^{13}$
cm.

The core collapse of the progenitor of 1993J resulted in the
deposition of about $3 \times 10^{50}$ erg of kinetic energy per solar
mass (Woosley et al. 1994). The expanded density profile shows an
outer steep power-law region with $n\sim 9$ and an inner, relatively
flat region (Blinnikov et al. 1998 and Fig. 4 therein). The bend in
the profile occurs at a velocity of $\sim 2000\;{\rm km\;s^{-1}}$. The
inner density profile is clearly not a single power law in
$r$. However, it is likely that hydrodynamic instabilities, similar to
those believed to occur in SN 1987A (e.g. Fryxell et al. 1991), lead
to mixing and smooth out any sharp features in the density
profile. Thus, a reasonable smooth, flat inner profile may be a good
approximation. In the outer parts of the density structure, on the
other hand, radiative transfer effects are more important for the
explosion of an extended star and can lead to the formation of a dense
shell in the outer layers. The shell is, however, expected out in the
steep power-law region of the density profile, where the pulsar bubble
is not likely to reach. In what follows, we assume $m=1$ or $2$ and
$n=9$.

The first stage of evolution involves the interaction of the pulsar
bubble with the inner density section. With the assumptions that the
pulsar luminosity is constant during this phase (i.e. $t_{\rm age} \le
t_\Omega$) and the supernova gas is swept up into a thin shell of mass
$M$ and velocity $v$, the radius of the pulsar bubble can be written
as (Chevalier \& Fransson 1992) \begin{equation} r_{\rm
p}=\left[{(3-m)(5-m)^3L_\Omega \over 4\pi A
(9-2m)(11-2m)}\right]^{1/(5-m)}t^{(6-m)/(5-m)}, \end{equation} where
$r_{\rm p} \propto v \propto M_{\rm t}^{-1/2} E^{(3-m)/2(5-m)}
L_\Omega^{1/(5-m)}$. The shell velocity is $v=$
$[740L_{\Omega,39}^{1/4}E_{51}^{1/4}M_{{\rm t},5}^{-1/2}t_{{\rm
age},11}^{1/4},270 L_{\Omega,39}^{1/3}E_{51}^{1/6}M_{{\rm
t},5}^{-1/2}t_{{\rm age},11}^{1/3}]$ $\;{\rm km\;s^{-1}}$ for
$m=[1,2]$, where $M_{{\rm t},5}=M_{\rm t}/ 5 M_\odot$,
$E_{51}=E/10^{51}$ erg, $L_{\Omega,39}=L_\Omega/10^{39}$ erg s$^{-1}$,
and $t_{{\rm age},11}=t_{\rm age}/ 11$ yr. The density transition
velocity is $v_{\rm t}=[5100,6300]E_{51}^{1/2}M_{{\rm t},5}^{-1/2}$
$\;{\rm km\;s^{-1}}$ for $m=[1,2]$, so it is likely that the shell
remains within the inner part of the density profile.

The density of the uniform gas shell is $\rho \approx [5 \times
10^{-19} L_{\Omega,39}^{-1/4}E_{51}^{-5/4}M_{{\rm t},5}^{5/2} t_{{\rm
age},11}^{-13/4}, 4 \times 10^{-15}
L_{\Omega,39}^{-2/3}E_{51}^{-5/6}M_{{\rm t},5}^{2} t_{{\rm
age},11}^{-11/3}]$ $\;{\rm g\; cm^{-3}}$ for $m=[1,2]$, where $\rho
\propto M_{\rm t}^{5/2}E^{-5(3-m)/2(5-m)}L_\Omega^{-m/(5-m)}$. The
presence of the companion implies that the supernova gas should be
transparent to optical radiation (i.e. the optical depth to electron
scattering should be less than unity), which gives $E > 10^{49}M_{{\rm
t},5}^{2}t_{{\rm age},11}^{-2}$ for $m=1$ and $L_\Omega \ge 10^{40}\;
E_{51}^{-2} M_{{\rm t},5}^{-9/2}\;t_{{\rm age},11}^{-7}$ $\; {\rm
erg\; s^{-1}}$ for $m=2$.

For X-ray energies $\ge$ 10 keV, electron scattering also provides the
main opacity. However, in this case the bound electrons contribute as
well as the free ones, so that the optical depth is always given by
electron scattering irrespectively of the degree of ionization. At
lower X-ray energies $\epsilon \ge$ 3 keV, photoionization provides
additional opacity (Bahcall et al. 1970) and may delay the time at
which the envelope becomes transparent by an additional factor $\sim
5(\epsilon/1{\rm keV})^{-3/2}$ (this estimate is for a predominantly
neutral envelope of pure hydrogen). The delay is somewhat larger if
the envelope is rich in heavy elements. The matter that is swept up by
the pulsar bubble is subject to Rayleigh-Taylor instability and may
form filaments (Vishniac 1983). The optical depth to electron
scattering could then be much less if the envelope is sufficiently
irregular that some lines of sight to the pulsar traversed relatively
little envelope mass. Under such favourable circumstances, the opacity
of the envelope is a less serious problem (except at late times when
grain formation may occur in the supernova ejecta and dust absorption
could significantly increase the opacity at visual wavelengths).

Radio measurements suggest that any pulsar nebula in the center of SN
1993J is fainter than $\approx 10^{38}$ erg s$^{-1}$ (Bietenholz et
al. 2003).  The material immediately surrounding the putative pulsar
is, however, expected to be totally opaque until
\begin{equation}
{t(\tau_{\rm ff}\sim 1)}=\left\{
\begin{array}{c@{\hspace{2em}}l}
2\times 10^{2}E_{51}^{-1/2}M_{{\rm t},5}T_4^{-3/4}\nu_9^{-1}\;{\rm yr} ~&~{\rm for}~\;m=1\\
10^{3}L_{\Omega,39}^{-1/7}E_{51}^{-2/7}M_{{\rm
t},5}^{-9/14}T_4^{-9/14}\nu_9^{-6/7}\;{\rm yr}~&~{\rm for}~\;m=2\\
\end{array}
\right.,
\end{equation}
at which time the radio flux ($\nu_9 = \nu/1$ GHz) will suddenly
appear. Here $T_4=T/10^4$K is the temperature of the swept-up
material. The lack of detection of a synchrotron nebula at radio
wavelengths should not be taken as strong evidence against the
presence of a bright pulsar (Bahcall et al. 1970). On the other hand,
the lack of an X-ray nebula indicates that, if one exists, its
radiative power must be less than $L_X \approx 2 \times 10^{38}$ erg
s$^{-1}$ (Zimmermann \& Aschenbach 2003).

\subsection{Emission Model}
A compelling model for the optical/X-ray properties of the Crab Nebula
was developed by Rees \& Gunn (1974). In this model, the central
pulsar generates a highly relativistic, particle dominated wind (with
Lorentz factor $\gamma_w$) that passes through a shock front and
decelerates to match the expansion velocity set by the outer
nebula. The emission from the pulsar bubble is thought to provide a
larger luminosity source than the radiative shock front itself
(Chevalier \& Fransson 1992). The wind particles acquire a power-law
energy spectrum of the form $N(\gamma) \propto \gamma^{-p}$ (for
$\gamma \ge \gamma_m$, where $\gamma$ is the particle Lorentz factor
and $\gamma_m$ is its minimum value) in the shock front and radiative
synchrotron emission in the down stream region (Chevalier 2000). Under
the basic simplification that the emitting region can be treated as a
one zone (i.e. no spatial structure in the nebula), and assuming that
a balance between injection from the shock front and synchrotron
losses is established, the number of radiating particles at a
particular $\gamma$ is given by $N(\gamma)=\gamma_m^{p-1} (\gamma_w
\beta B^2)^{-1}L_\Omega\gamma^{-(p+1)}$ (Chevalier 2000), where $B$ is
the magnetic field in the emitting region, and $\beta=1.06 \times
10^{-15}$ cm$^{3}$ s$^{-1}$.

In what follows, it is assumed that the energy density in the emitting
region (which is approximately determined by the shock jump
conditions) is divided between a fraction $\epsilon_e$ in particles
and a fraction $\epsilon_B$ in the magnetic field. These efficiency
factors are constrained by $ \epsilon_{e}+ \epsilon_{B}=1$. With this
assumption, the magnetic field in the emitting region is
$B=(6\epsilon_B L_\Omega/r_s^2c)^{1/2}$, where $r_s$ is the shock wave
radius. The electron energy is radiated at its critical frequency
$\nu(\gamma)=\gamma^2 (q_eB/2\pi m_e c)$, where $q_e$ and $m_e$ are
the electron charge and mass, respectively. If the electrons and
positrons cool rapidly by synchrotron radiation, as thought to be the
case for the Crab Nebula, the luminosity produced from the pulsar
power is given by $L_\nu =\phi(p) \epsilon_e^{p-1}
\epsilon_B^{(p-2)/4} \gamma_w^{p-2} r_s^{2-p)/4}
L_\Omega^{(p+2)/4}\nu^{-p/2}$ (Chevalier 2000), where $\phi(p)={1\over
2}({p-2 \over p-1})^{p-1}({6\psi^2 \over c})^{(p-2)/4}$ and
$\psi=2.8\times 10^6$ in cgs units. If the particle spectrum is
similar to that of the Crab Nebula ($p=2.2$), the X-ray luminosity
($\nu=10^{18}$ Hz) becomes
\begin{equation}
\nu L_\nu \approx 5 \times 10^{37} \epsilon_{e}^{6/5}
\epsilon_{B}^{1/20} \gamma_{w,6}^{1/5} r_{s,16}^{-1/10}
L_{\Omega,39}^{21/20} \nu_{18}^{-1/10} {\rm erg\;s^{-1}},
\label{Lx}
\end{equation}
where cgs units are used. We note that the X-rays from a putative
pulsar nebula in SN 1993J with $L_{\Omega} \sim 10^{39}\;{\rm
erg\;s^{-1}}$ could provide the additional energy input required to
reproduce the observed H$\alpha$ luminosity at $\sim$350 days (Houck
\& Fransson 1996).

The above relation can be used to predict the spin-down power,
$L_\Omega$, of the putative pulsar in wind nebula where a pulsar has
not yet been observed (Chevalier 2000). In the case of SN 1993J, the
value $r_s$ can be determined by the condition that the envelope
should be transparent to optical radiation (Section 2.1). The need for
a particle dominated shock suggests that $\epsilon_{e}\ge 0.5$. We set
$\epsilon_{B}=\epsilon_{e}=0.5$, although there is little dependence
to $\epsilon_{B}$. The value of $\gamma_w$ is difficult to
constraint. Here we use $\gamma_w \sim 3\times 10^6$, the value
typically assumed for the Crab Nebula (Kennel \& Coroniti 1984). A
better estimate of $\gamma_w$ is given in Section 3. When these
parameters are substituted into equation (\ref{Lx}), the {\it XMM}
upper limit of $L_X \approx 2 \times 10^{38} {\rm erg\;s^{-1}}$
(0.3-10 keV; Zimmermann \& Aschenbach 2003) yields $L_\Omega \le 2
\times 10^{39} {\rm erg\;s^{-1}}$.  On the basis of the {\it ASCA}
data, Kawai et al. (1998) found the relation $\log L_X = (33.42 \pm
0.20)+(1.27\pm0.17)\log(L_\Omega/10^{36})$, where $L_X$ is the nebular
luminosity in the 1-10 keV range, from which we estimate $L_\Omega \le
5 \times 10^{39} {\rm erg\;s^{-1}}$. This estimate is in fair
agreement with our predicted value.

\section{The Role of Binarity}
At the time of the explosion, the progenitor of 1993J has a mass of
5.4 $M_\odot$ (with a helium-exhausted core of 5.1 $M_\odot$), the
secondary has a mass of 22 $M_\odot$. The orbital period of the system
is $\sim$ 25 yr, and the companion has an orbital velocity of $\sim$ 6
km s$^{-1}$ (Maund et al. 2004). There may have been a large kick
imparted by the explosion mechanism, and that would be very
interesting to study, but baring that, let us assume the companion
star is bound in an eccentric orbit with the newly born
pulsar\footnote{The binary is likely to remain bound after the
  supernova explosion provided that the kick velocity $\la$ 90
  ${\rm\;km\;s^{-1}}$ (Brandt \& Podsiadlowski 1995).}. The bulk of
the pulsar energy $L_\Omega$ would be primarily in the form of a
magnetically driven, highly-relativistic wind consisting of $e^{-}$,
$e^+$ and probably heavy ions with $L_w \simeq \zeta L_\Omega$ and
$\zeta \le 1$.

Under the foregoing conditions, the relativistic wind (which is likely
to be undisturbed by the presence of the binary
companion\footnote{Here $v_{\rm w}$ and $\dot{M}_{\rm w}$ are the
velocity and mass-loss rate of the stellar companion.} since $L_\Omega
\gg v_{\rm w}^2\dot{M}_{\rm w}$) would escape the compact remnant
while interacting with the soft photon field of the companion with
typical energy $\theta_\ast=kT_\ast/ (m_ec^2) \sim 3\times 10^{-6}$
(Maund et al. 2004). The scattered photons whose energy is boosted by
the square of the bulk Lorentz factor of the magnetized wind
(i.e. $\Theta_*=2\gamma_w^2 \theta_* \sim 6 \times 10^{6}
\gamma_{w,6}^2$) propagate in a narrow $\gamma_w^{-1}$ beam owing to
relativistic aberration. The rate of energy loss of a relativistic
particle moving in a radiation field with an energy density $w_*
\simeq L_*/ 4\pi a^2 c$ is about $m_e c^2 d \gamma/dt \approx -
w_*\sigma_T c \gamma^2$ in the Thompson limit (Landau \& Lifshitz
1975). Here $L_*\sim 10^5 L_\odot$ is the luminosity of the optical
star and $a$ is the binary separation ($a\sim 25\;{\rm AU} \sim
200R_\ast$; Maund et al. 2004). The total luminosity of scattered hard
photons $L_\Gamma$ is then equal to the total particle energy losses
in the course of motion from the pulsar to infinity $L_\Gamma= L_w
{\Delta \gamma \over \gamma} \simeq 1-(1+ \varsigma)^{-1}$, where
\begin{equation}
\varsigma={\gamma_w \sigma_T L_\ast \over 4\pi a m_ec^3} \sim
\left({\gamma_w \over 3 \times 10^6}\right) \left({L_\ast \over 10^5
L_\odot}\right) \left({a \over 200 R_*}\right)^{-1}
\label{eff}
\end{equation}
denotes the efficiency in extracting energy from the relativistic
outflow (Chernyakova \& Illarionov 1999). The resulting radiation
pressure on electrons in the ejecta will brake any outflow whose
initial Lorentz factor exceeds some critical value $\gamma_{\rm lim}
\leq {a \over R_*} ({L_\Omega \over L_*})^{1/2} \sim 3 \times 10^2
L_{\Omega,39}^{1/2}$, converting the excess kinetic energy into a
directed beamed of scattered photons. With $\gamma_w \sim \gamma_{\rm
lim}$, equation (\ref{Lx}) yields $L_\Omega \le 6 \times 10^{39} {\rm
erg\;s^{-1}}$.

\begin{figure*}
\centerline{\psfig{file=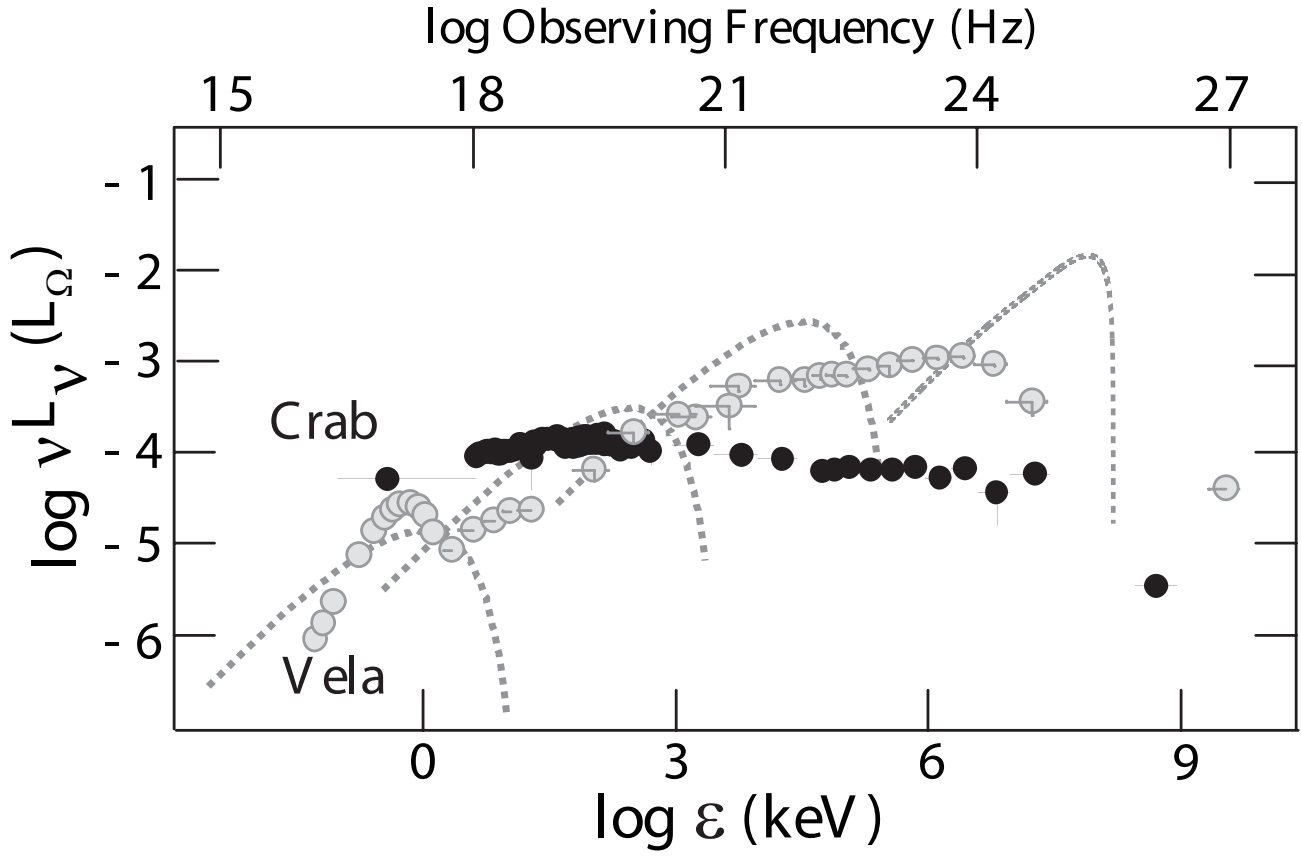,width=\textwidth}}
\caption{Representative spectra produced by the interaction of the
pulsar wind with the radiation of the stellar companion. The
luminosity of the gamma-ray emission $L_\Gamma$ is shown for
$\gamma_w=10^2,\;10^3,\;10^4,\;10^6$ (dashed lines). The spectra of
the Crab and Vela pulsars are shown for comparison (adapted from
Thomson 2003).}
\label{fig1} 
\end{figure*}

As the stellar companion emits a black body spectrum, of effective
temperature $\theta_\ast$, the local photon energy density is given by
\begin{equation}
n(r,\varpi)={2\pi \over h c^3}\left({m_e c^2 \over h}\right)^2
\left({R_* \over a }\right)^2 {\varpi^2 \over
\exp[\varpi/\theta_\ast]-1},
\label{den}
\end{equation}
where $\varpi$ is the soft photon energy in units of $m_ec^2$. The
scattered photons are boosted by the square of the Lorentz factor so
that the local spectrum has a black body shape enhanced by
$\gamma_w^2$. As can be seen in Fig. \ref{fig1}, the resulting
spectrum is the convolution of all the locally emitted spectra
(i.e. $\int_0^{\infty}\;dn_\varpi[r,\varpi]$) and it is not one of a
blackbody. Note that Klein-Nishina effects are important for incoming
photon energies such that $\theta_\ast \gamma_w \ga 1$.  The maximum
energy of the scattered photons in this regime is $\gamma_w m_ec^2$.

We note that the total luminosity emitted by the relativistic
outflowing wind through the Compton-drag process in the direction of
the observer could be highly anisotropic and may change periodically
during orbital motion for an eccentric (or a highly inclined)
orbit. The time dependence, in this case, will be very distinctive and
such effects should certainly be looked for.

\section{Discussion}
\begin{figure*}
\centerline{\psfig{file=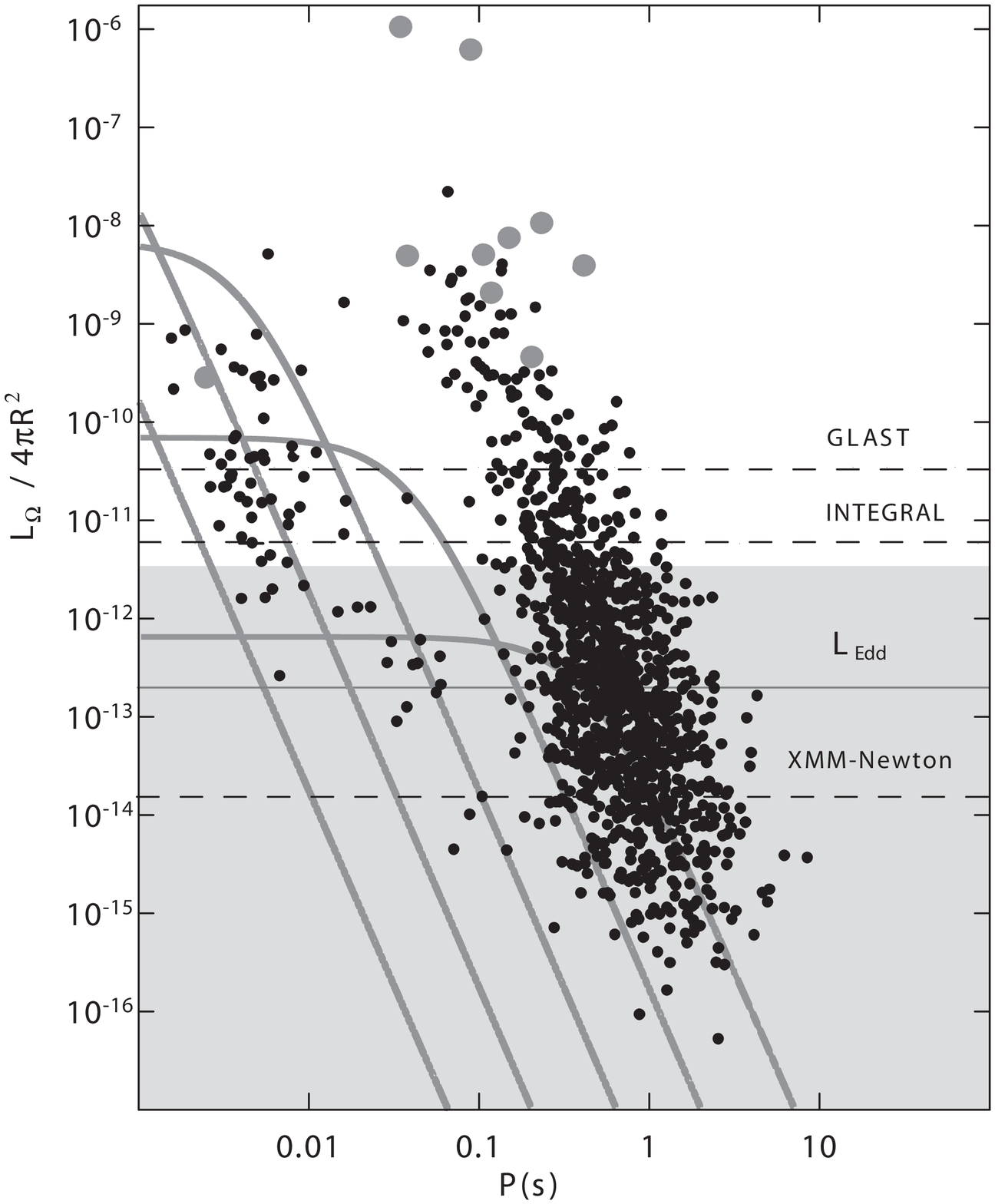,width=0.8\textwidth}}
\caption{Pulsar observability as measured by the spin-down energy seen
at Earth. Small dots represent pulsars with no gamma-ray emission,
while large filled symbols show all known gamma-ray pulsars (including
candidates; Thompson 2003).  The power-output of the pulsar wind, at
the distance and age (here assumed to be 11 years) of SN 1993J, is
shown for $B=10^{11},\;10^{12},\;10^{13},\;10^{14}$ and $10^{15}$ G
(from left to right) as a function of the assumed initial period. The
putative pulsar in SN 1993J, spinning down by magnetic dipole
radiation alone, can have a wind nebula X-ray luminosity below the
{\it XMM} limit if $L_\Omega \le 6 \times 10^{39} {\rm erg\;s^{-1}}$
(shaded region).}
\label{fig2} 
\end{figure*}

Despite presumptions that a neutron star may have been created when
the progenitor star of SN 1993J exploded and its core collapsed, no
pulsar has yet been seen. If one exists, its radiative power must be
less than $L_\Omega \approx 6 \times 10^{39} {\rm erg\;s^{-1}}$ or
lower if the pulsar nebula has a high radiative efficiency. The
well-known Crab pulsar and its nebula for comparison, put out $2
\times 10^{38}$ erg s$^{-1}$; and originally, when it was spinning
faster, the luminosity might have been seven times greater
still. Fig. \ref{fig2} shows the pulsar spin-down luminosity divided
by $4\pi$ times the square of the distance, the total pulsar energy
output at Earth. The ten gamma-ray pulsars (including candidates) are
shown as large circles.  The solid curves show the dipole emission of
the putative pulsar in 1993J (at $t\sim 11$ years) for various
assumptions regarding its initial period and magnetic field
strength. The pulsar spinning down by magnetic dipole radiation alone,
with initial rotation periods of 10-30 ms -- as extrapolated for
galactic young pulsars -- can have a spin-down luminosity below $\sim
6 \times 10^{39} {\rm erg\;s^{-1}}$ (see shaded region in
Fig. \ref{fig2}) only if either the magnetic field is relatively weak
$\le 10^{11}$ G or if it is so strong (i.e. $\ge 10^{15}$ G) that the
pulsar luminosity decays rapidly. If weak magnetic moments could be
ruled out in the near future by X-ray and infrared observations, then
we find that, if undetected, the putative pulsar in SN 1993J could be
a magnetar.

With a spectrum similar to that of the Crab (see Fig. \ref{fig2}), the
pulsed emission is likely to be below the detection limit of current
instruments and may deprive us of the opportunity to witness the
emergence of a gamma-ray pulsar in the immediate future. GeV radiation
with luminosities high enough to be detected with {\it
ARGO}\footnote{http://argo.na.infn.it/}, and the {\it
Veritas}\footnote{http://veritas.sao.arizona.edu/} experiment now
under construction, could be produced by the interaction of the pulsar
wind with the soft photon field of its companion provided that the
system remains bound and $\gamma_w \ge 10^6$ (i.e. similar or larger
to than inferred for the Crab pulsar). Its detection will surely offer
important clues for identifying the nature of the progenitor and
possibly constraining whether or not kicks have played an important
role.

If there is not a neutron star in SN 1993J, could there be a black
hole instead? Had matter fallen back onto the nascent neutron star
(with mass of 1.4 $M_\odot$) on a timescale of 100 seconds to a few
hours after the explosion, enough mass may have been accreted to push
the object over the minimum thought to be necessary for the creation
of a black hole. Matter would continue to accrete onto the black hole,
but the resulting radiation would be trapped in the outflow so that
the escaping luminosity would be small. Because the cores of massive
stars increase with initial stellar mass, this picture would be more
plausible if the initial mass of the SN 1993J progenitor star,
$\sim$15 $M_\odot$, were close to the mass limit above which stars
collapse directly to black holes (e.g. Fryer 1999).

Even if the neutron star is rotating only slowly or has a weak
magnetic field, one would expect surrounding material to fall back
onto it, giving rise to a luminosity $\lse L_{\rm Edd} \sim 3 \times
10^{38}$ erg s$^{-1}$ (see Fig. 1). The remaining possibility of a
weak pulsar with little surrounding mass will be difficult to rule
out. The thermal emission from a newly formed neutron star gives a
luminosity of $\sim 3 \times 10^{34}$ erg s$^{-1}$. The neutron-star
emission should have a characteristic soft X-ray emission, but its
detection will be arduous. If future observations fail to identify a
pulsar in 1993J, then the youngest pulsar that we know will still be
PSR J0205+6449 (Camilo et al. 2002), associated with SN 1181.

\section*{Acknowledgements}
This work was stimulated by conversations with M.~J. Rees and
S. Smartt. Discussions with J. Bahcall and R. Chevalier are gratefully
acknowledged, as is helpful correspondence from the referee
Ph. Podsiadlowski. This work is supported by the W.M. Keck foundation
(AMS) and NASA through a Chandra Postdoctoral Fellowship award
PF3-40028 (ER-R).

\clearpage

\begin{thebibliography}{}

\bibitem{} Aldering G., Humphreys R.~M., Richmond M., 1994, AJ, 107,
662

\bibitem{} Arnett W.~D., 1988, ApJ, 331, 377

\bibitem{} Bahcall J., Rees M.~J., Salpeter E.~E., 1970, ApJ, 162, 737

\bibitem{} Bietenholz M.~F., Bartel N., Rupen M.~P., 2003, ApJ, 597,
374
\bibitem{} Blinnikov S.~I., Eastman R., Bartunov O.~S., Popolitov
V.~A., Woosley S.~E., 1998, ApJ, 496, 454

\bibitem{} Brandt N., Podsiadlowski Ph., 1995, MNRAS, 274, 461

\bibitem{} Camilo F. et al., 2002, ApJ, 571, L41

\bibitem{} Chernyakova M.~A., Illarionov A.~F., 1999, MNRAS, 304, 359

\bibitem{} Chevalier R.~A., 1997, Science, 276, 1374

\bibitem{} Chevalier R.~A., Soker N., 1989, ApJ, 341, 867

\bibitem{} Chevalier R.~A., Fransson R.~A., 1992, ApJ, 395, 540

\bibitem{} Chevalier R.~A., 2000, ApJ, 539, L45

\bibitem{} Filippenko A.~V., Matheson T., Ho L.~C., 1993, ApJ, 415,
L103

\bibitem{} Freedman W.~L. et al., 1994, ApJ, 427, 628

\bibitem{} Fryer C., 1999, ApJ, 522, 413

\bibitem{} Fryxell B., Muller E., Arnett W.~D., 1991, ApJ, 367, 619

\bibitem{} Houck J.~C., Fransson R.~A., 1996, ApJ, 456, 811

\bibitem{} Kawai N., Tamura K., Shibata S., 1998, in Neutron Stars and
Pulsars; Thirty Yeras after the Discovery, ed. N.Shibazaki, N. Kawai,
S. Shibata, \& T. Kifune (Tokyo; Universal Academy), 449


\bibitem{} Kennel C.~F., Coroniti F.~V., 1984, ApJ, 283, 694

\bibitem{} Landau L.~D., Lifshitz E.~M., 1975, The Classical Theory of
Fields. Pergamon Press, New York

\bibitem{} Maund J.~R. et al., 2004, Nature, 427, 129

\bibitem{} Nomoto K. et al., 1993, Nature, 364, 507

\bibitem{} Podsiadlowski Ph., Hsu J., Joss P., Ross R.~R., 1993,
Nature, 364, 509

\bibitem{} Rees M.~J., Gunn J.~E., 1974, MNRAS, 167, 1

\bibitem{} Rickett B.~J., Seiradakis J.~H., 1982, ApJ, 256, 612

\bibitem{} Shapiro S.~L., Teukolsky S.~A., 1983, Black Holes, White
Dwarfs, and Neutron Stars: The Physics of Compact Objects. Wiley, New
York

\bibitem{} Shigeyama T., Nomoto K., 1990, ApJ, 360, 242
 
\bibitem{} Thompson J.~T., 2003, astro-ph/0312272

\bibitem{} Vishniac E.~T., 1983, ApJ, 274, 152

\bibitem{} Woosley S.~E., Eastman R.~G., Weaver T.~A., Pinto
P.~A., 1994, ApJ, 429, 300

\bibitem{} Zimmermann H.~U., Aschenbach B., 2003, A\&A, 406, 969


\end{thebibliography}
\end{document}